# Cosmic Microwave Background Anisotropies from the Rees-Sciama Effect in $\Omega_0 \leq 1$ Universes

Robin Tuluie[1] and Pablo Laguna[2]

*Department of Astronomy and Astrophysics*
*and Center for Gravitational Physics and Geometry*
*The Pennsylvania State University, University Park, PA 16802*

and

Peter Anninos[3]

*Laboratory for Computational Astrophysics*
*National Center for Supercomputing Applications*
*University of Illinois at Urbana-Champaign*
*Beckman Institute, 405 N. Mathews Ave.*
*Urbana, IL 61801*

## ABSTRACT

We investigate the imprint of nonlinear matter condensations on the Cosmic Microwave Background (CMB) in $\Omega_0 < 1$ cold dark matter (CDM) model universes. We consider simulation domains ranging from $120h^{-1}$ Mpc to $360h^{-1}$ Mpc in size. We concentrate on the secondary temperature anisotropies induced by time varying gravitational potentials occurring after decoupling. Specifically, we investigate the importance of the Rees-Sciama effect due to: (1) intrinsic changes in the gravitational potential of forming, nonlinear structures, (2) proper motion of nonlinear structures, and (3) late time decay of gravitational potential perturbations in open universes. CMB temperature anisotropies are obtained by numerically evolving matter inhomogeneities and CMB photons from an early, linear epoch ($z = 100$) to the present, nonlinear epoch ($z = 0$). We test the dependence and relative importance of these secondary temperature anisotropies as a function of the scale of the underlying matter (voids, superclusters) and as a function of $\Omega_0$. The results of the $\Omega_0 < 1$ models are compared to a similarly executed $\Omega_0 = 1.0$ simulation. We find that in low density models all three sources of anisotropy could be relevant and reach levels of $\Delta T/T \sim 10^{-6}$. In particular, we find that for $\Omega_0 < 1$ at large scales, secondary temperature anisotropies are dominated by the decaying potential.

---

[1] rtuluie@astro.psu.edu

[2] pablo@astro.psu.edu

[3] panninos@ncsa.uiuc.edu

*Subject headings:* cosmic microwave background — cosmology: theory

## 1. Introduction

The cosmic microwave background radiation is a direct relic of the early universe. It is a unique and deep probe of both the thermal history of the early universe and the primordial perturbations in the matter distribution. However, the exclusively primordial origin of anisotropies in the CMB is not guaranteed. Important modifications to the CMB spectrum and level of anisotropy can occur even well after decoupling. If a substantial late reionization of the intergalactic medium (IGM) occurs sufficiently early, the increased rate of Thomson scattering will erase sub-horizon scale temperature anisotropies, while recreating secondary Doppler shift temperature anisotropies at these and smaller scales. (Vishniac 1987, Tuluie et al. 1995, Dodelson & Jubas 1993). In this paper we will focus on another source of secondary temperature anisotropy from late-time gravitational effects. One would expect the CMB to carry a signature of the evolving matter structures of the universe through the gravitational red and blue shifting of the CMB photons as they traverse the evolving gravitational potential wells of the forming matter structures. This additional source of anisotropy therefore is due to non-static gravitational potentials. In the case of a flat background universe ($\Omega_0 = 1$), gravitational potential perturbations are static in the linear regime of matter evolution ($\delta\rho/\rho \ll 1$). However, in the nonlinear regime and quasi-linear regime this is no longer the case; the evolution of gravitational potential wells will leave an imprint on the CMB. In addition, if the universe is open ($\Omega_0 < 1$), the gravitational potential perturbations will decay even while the matter structures are still in the linear regime (Mukhanov et al. 1992). This decay occurs independently of the clustering matter structures and presents an additional source of anisotropy in the microwave sky.

We categorize these secondary anisotropies induced by time-varying potentials along the photon trajectories into three classes: (1) *the intrinsic Rees-Sciama effect* due to evolving gravitational potentials of nonlinear and quasi-linear matter structures as a result of collapse and expansion (Rees & Sciama 1968); (2) *the proper motion effect* caused by proper motion of clusters and superclusters across the microwave sky (Birkinshaw & Gull 1983, Stebbins 1987, Tuluie & Laguna 1995, hereafter TL95) and (3) the *decaying potential effect* from the decay of gravitational potential perturbations due to the rapidly expanding background of the $\Omega_0 < 1$ models (Kamionkowski & Spergel 1994).

It is important to accurately understand sources of secondary perturbations to the CMB; many linear analysis are based on the assumption that some secondary nonlinear sources, such as the Rees-Sciama effect, are negligible (for a comprehensive review, see White et al. 1994). An affirmation of this assumption for the entire parameter space allowed in current dark matter models will offer reassurance to the majority of such CMB spectrum calculations, while the result of a large secondary source of anisotropy would require a more computationally involved technique such as ours. Especially in light of the attempts to obtain $\Omega_0, \Omega_b$ and $h$ from the location, width and height of the Doppler peaks in the CMB spectrum (White et al. 1994 and references therein), it is important to assure that these analyses did not make unjustified assumptions about the importance of secondary nonlinear perturbations.

The total secondary CMB anisotropy, in addition to the relative level of these three secondary sources, depends on both the parameters of the background model (i.e. $\Omega_0$) and the particular clustering level and size of the nonlinear and quasi-linear matter structures. In this paper, we investigate the relative and absolute contributions of these three sources of anisotropy to the primordial Sachs-Wolfe effect (Sachs & Wolfe 1967). We expand the treatment of TL95 to include larger structures, separate the decaying potential effect from the total Rees-Sciama effect, study the correlation between the sources of secondary gravitational anisotropy and the underlying density field and potential wells, and investigate the relative importance of these three sources of secondary CMB anisotropy as a function of void scale, angular scale and $\Omega_0$. We also include the Einstein-deSitter $\Omega_0 = 1$ case for comparison. For this purpose, we have ignored Thomson scattering for epochs $z \ll z_{dec} \sim 1100$, as is the case for no reionization of the IGM. Reionization is discussed in Tuluie et al. 1995 and would have little effect on these late ($z \sim 10$ or less) anisotropies.

The paper is organized as follows. Section 2 describes the three sources of secondary, gravitationally induced temperature anisotropies: the intrinsic Rees-Sciama, the proper motion and the decaying potential effects. Section 3 discusses the matter simulation which provides the backdrop against which we wish to compute temperature anisotropies. After a brief review of the current status of the observed large scale structure of the universe, we discuss the matter model's normalization, initialization and evolution. Our ray tracing technique is used in section 4 to obtain temperature anisotropies from the matter model of section 3. First we describe the technique and test it's accuracy. Next we separate some of the three sources of secondary temperature anisotropy out of our skymaps in order to ascertain their individual importance. Results are presented in section 5 and conclusions in section 6.

## 2. Secondary Temperature Anisotropies from Gravitational Effects

### 2.1. The Rees-Sciama Effect

The Rees-Sciama effect creates additional temperature anisotropies in CMB photons which freely traverse the universe. Specifically, its source is the changing depth of the gravitational potential wells of collapsing clusters and superclusters and the nonstatic potential perturbations of expanding and evacuating voids. Photons, as they propagate along their geodesics through these evolving structures, are subjected to the gravitational blue and red shifting as they enter and leave structures. If the photon crossing time for a particular evolving structure is an appreciable fraction of the structure evolution time scale, the Rees-Sciama effect may be a significant fraction of the Sachs-Wolfe anisotropy.

During the linear epoch of cosmological perturbations, that is when perturbations satisfy both $\delta\rho/\rho \ll 1$ and $\phi \ll 1$, the evolution of the density field is governed by a gravitational field $\phi$ which represents small perturbations against the homogeneous background. In a $\Omega_0 = 1$ universe, the gravitational potential $\phi$ is static as long as the density perturbations are linear. The nonlinear epoch is reached once the density perturbation field $\delta\rho/\rho$ has grown to values of order unity. However, even though the density perturbation field may now be nonlinear, $\delta\rho/\rho \geq 1$, the underlying gravitational field is still well described by linear theory; that is, first order gravitational perturbation theory with $\phi \ll 1$ holds. On sub-horizon scales this reduces to Newtonian gravity in comoving coordinates (Mukhanov et al. 1992). Once the nonlinear epoch is reached, the gravitational field can become significantly nonstatic due to the evolution of the density field. CMB photons, which mirror the gravitational field $\phi$ can thus reflect a signature of the nonlinear density perturbation field, in addition to the usual imprint of the gravitational potentials at the surface of last scattering. If $\Omega_0 < 1$, the underlying gravitational field is not static. Instead, once the background evolution becomes curvature dominated, the gravitational perturbations $\phi$ decay (Mukhanov et al. 1992). This decay is again mirrored in the CMB, which is now expected to carry the additional signature of the decaying gravitational field.

The temperature anisotropies of the CMB associated to gravitational potentials and proper motions are given by (Sachs & Wolfe 1967)

$$\frac{\Delta T}{T} = \phi_r - \phi_e + 2\int_r^e \frac{\partial \phi}{\partial t} dt, \qquad (1)$$

where the first term represents the Sachs-Wolfe effect and the second term is the total secondary gravitational temperature anisotropy. The numerical factors multiplying the Sachs-Wolfe and integral term in equation (1) differ among various authors depending on the choice of gauge (Padmanabhan 1993). Our results were obtained using the longitudinal gauge only.

Many analytic and some numerical calculations of anisotropies of the CMB that are due to the late time evolution of the gravitational potential perturbations have been presented in the literature (Quilis et al. 1995, Seljak 1995, Vadas 1994, Martinez-González et al. 1994, Hu & Sugiyama, Meszáros 1994, Panek 1992, Chodorowski 1991, Dyer & Ip, Thompson & Vishniac 1987, Kaiser 1982, Rees & Sciama 1968). However, most of these studies have considered only the effect of a single, spherically symmetric density inhomogeneity on the CMB anisotropy using either a Swiss cheese model, thin-shell approximation or Tolman-Bondi solution. At best, the analytical models only roughly describe the correct representation and late time evolution of the evolving inhomogeneities. The effect on the CMB of multiple, non-trivially distributed matter structures through which the CMB photons propagate, as well as the complete and unambiguous continuous linkage between the linear, weakly clustered and fully nonlinear epochs, has important consequences (TL95). In addition, the peculiar transverse bulk motion of the matter structures across the microwave sky has to be taken into account and provides a significant contribution to the Rees-Sciama effect at angular scales of a few degrees (TL95).

For $\Omega_0 = 1$ the Rees-Sciama effect is expected to scale with structure and void size as follows (Panek 1992, see also Thompson & Vishniac 1987 and Meśzaŕos 1994):

$$\Delta T/T \sim \left(\frac{\delta\rho}{\rho}\right)^{3/2} \left(\frac{d}{t}\right)^3, \qquad (2)$$

where $d$ is the physical size of the structure and $t$ is the structure crossing time of the CMB photons. However, this simple scaling law may lead to a weaker dependence on void scale even if very large voids are juxtaposed in an $\Omega_0 = 1$ universe if the void density contrast decreases sufficiently for large voids. In particular, for the scale-invariant Harrison-Zel'dovich spectrum, the large scale density contrasts scale with comoving wavelength $\lambda = d/a(t)$ as $(\delta\rho/\rho)^2 \propto \lambda^{-4}$, so $\Delta T/T \propto (\lambda^{-2})^{3/2} (d/t)^3 \sim$ const., which leads to an approximately constant dependence of the Rees-Sciama effect with void size for the Harrison-Zel'dovich spectrum.

It is clear that whatever the approach is to obtain accurate estimates of the Rees-Sciama effect, a realistic representation of the underlying matter structure is essential. In this respect, the careful normalization of a computer simulation is as important as the use of reasonable and observationally plausible values for structure sizes *and* density contrasts in analytic estimates. In addition, it is important to expand the numerical simulations to include scales and void sizes which are large, even by the standards of the largest observed void sizes, like the $60h^{-1}$ Mpc Bootes void. However, even if these large scale voids do exist, their density contrast is likely to be bound by the Harrison-Zel'dovich spectrum (Scaramella 1992).

### 2.2. The Proper Motion Effect

In previous work (TL95), we have analyzed the Rees-Sciama effect for a $\Omega_0 = 1$ CDM universe. We have found that a major contribution to the Rees-Sciama effect comes from the overdense structures (see fig. 3 of TL95) and that the signature of typical $\sim 30h^{-1}$ Mpc voids is small. We also found that the contributions from positive density enhancements, such as clusters and superclusters, is comprised of both the intrinsic effect due to the collapse of the forming cluster or supercluster, and a signature whose source is the bulk motion of the cluster or supercluster across the microwave sky.

A moving cluster or supercluster creates a time-changing potential perturbation and can therefore leave an imprint in the CMB (Birkinshaw & Gull 1983, Stebbins 1987). Fig. 1 shows the dipolar imprint left in the CMB sky by a transversely moving, but intrinsically static, spherical cluster. Specifically, if the cluster moves transversely to the direction of observation, CMB photons which enter the potential perturbation of the cluster ahead of its path will be redshifted while photons entering the cluster's wake will be blueshifted. A CMB photon entering the wake of an intrinsically static but transversely moving cluster potential will be traveling through a potential well which is decreasing in overall depth as the photon crosses it. The potential perturbation and potential gradients will be smaller upon the photon's exit than during its entry into the moving cluster's potential. This photon will be blueshifted more upon entering the potential than it will be redshifted during its exit and a net blueshift with respect to a background photon results. Similarly, a photon entering and exiting the cluster's potential well ahead of the approaching cluster will be redshifted as the potential well depth increases during the photon crossing time.

A dipolar imprint, with the center between the hot (blue) and cold (red) lobes located at the cluster's center appears in the CMB. This proper motion effect, in particular the temperature difference between the dipole maximum and minimum temperatures, scales proportional to the depth of the cluster potential well and the transverse velocity of the moving cluster. The angular separation of the maximum and minimum temperatures of the dipolar lobes is a measure of the extend of the cluster's potential well. Furthermore, one can in principle use the proper motion effect to measure the transverse velocity of the cluster or obtain further information of its potential well and hence its dark matter content and biasing. The maximum proper motion effect for a single cluster or supercluster occurs for purely transverse motion and is given by (Birkinshaw & Gull 1983):

$$\Delta T/T \sim 1.5 \times 10^{-7}(v/1000 \; km \; s^{-1})(M_0/10^{15}M_\odot)(R_0/10Mpc)^{-1}, \qquad (3)$$

where $M_0$ is the mass of the cluster distributed in a constant spheroid of radius $R_0$. The result for a spherical density profile of the form $\rho(r) = \rho_0(1+(r/r_{core})^2)^{-\xi}$ is of similar magnitude (Stebbins 1987).

We attempt to separate the intrinsic Rees-Sciama and proper motion effect from the decaying potential effect for our models in section 4.2.

### 2.3. The Decay of Potential Perturbations in the Open Model

If the universe is open ($\Omega_0 < 1$), the gravitational potential perturbations will no longer remain static during the linear phase of structure formation. Instead, the increased rate of expansion of the background reduces the amplitude of potential perturbations. The decay of the potential fluctuations will be mirrored in the CMB photons. This decay occurs independently of the clustering matter structures and presents an additional source of anisotropy in the microwave sky.

The evolution of the potential fluctuations is obtained by explicitly integrating the equation for the gauge invariant density perturbations. For the flat ($\Omega_0 = 1$), matter dominated model the gravitational potential is given by

$$\phi(\mathbf{x},a) = C_1(\mathbf{x}) + C_2(\mathbf{x})a^{-5/2}. \qquad (4)$$

with $a(t) \propto t^{2/3}$ and $C_1(\mathbf{x})$ and $C_2(\mathbf{x})$ describing the spatial dependence of the potential. Dropping the decaying mode term in equation (4), the gravitational potential remains static for a flat universe.

For $\Omega_0 < 1$ an analytic expression for the growing mode solution for $\phi(\mathbf{x}, a)$ is given as

$$\phi(\mathbf{x}, a) = C_1(\mathbf{x}) \left(\frac{a}{a_m}\right)^{-3} \left[2\frac{a^2}{a_m^2} + 12\frac{a}{a_m} - 6\sqrt{\frac{a^2}{a_m^2} + 2\frac{a}{a_m}} \ \ln\left(\frac{a}{a_m} + 1 + \sqrt{\frac{a^2}{a_m^2} + 2\frac{a}{a_m}}\right)\right], \quad (5)$$

where $a_m = \Omega_0/(2(1 - \Omega_0))$. This analytic solution is shown in figure 2 for $\Omega_0 = 0.3$. The potential decay occurs as early as $z \approx 20$, at which point a 10% reduction in the original amplitude of $\phi$ is reached. By $z = 0$, $\phi$ has been reduced by a factor of two from its primordial value.

We expect this to be a substantial source of anisotropy contributing to the primordial Sachs-Wolfe effect. However, as CMB photons traverse regions of both positive and negative potentials and their associated potential decay, cancelation between these secondary induced anisotropies will reduce their overall effect. These results are presented in section 5.

## 3. Matter Model

The problem of the level of temperature fluctuations of the CMB is closely tied to the problem of structure formation. In order to ascertain CMB fluctuation levels in a particular realization of the universe, be it by computer simulations or by purely analytical considerations, one has to solve the problem of structure formation first.

Many surveys of the large scale structure of the universe have been carried out (Bahcall and Soneira 1983, De Lapparent et al. 1986, Bahcall 1988, Geller and Huchra 1986, Maddox et al. 1990, Broadhurst et al. 1990, Loveday et al. 1992) which provide extensive evidence of clustering on scales of up to $\sim 150h^{-1}$ Mpc. In particular, these surveys indicate a filament/sheet/knot structure of the large scale matter condensations, with some objects, such as the "Great Wall", extending over distances of $\sim 170h^{-1}$ Mpc. It is found that galaxies cluster with a slope $\gamma = -1.8$ of the correlation function $\xi_{gg}(r) = (r/r_o)^\gamma$ and that the correlation function reaches unity on scales of $r_o = 5.4h^{-1}$ Mpc (Peebles 1993). For rich clusters of galaxies the correlation function $\xi_{cc}(r)$ has been determined to have a slope of $\gamma = -1.8$ and reaches unity at $r_o = (20 - 25)h^{-1}$ Mpc (Bahcall 1988). The inhomogeneous distribution of such clusters can lead to density contrasts of $\delta\rho/\rho \geq 1$ on scales up to $25h^{-1}$ Mpc. These measurements form a solid yardstick against which any scenario of structure formation must be measured.

Besides the highly clustered large scale structure of the universe, there is evidence for large scale coherent velocity flows of the gravitating matter (Dressler et al. 1987, Bertschinger et al. 1990). The detection of bulk velocity flows on large scales gives credence to the "Great Attractor" model of Lynden-Bell et al. (1988) in which peculiar motions of clusters of galaxies are dominated by large coherent flows over large regions. The observations have found extensive coherent velocity flows on scales up to $\sim 100 h^{-1}$ Mpc with streaming velocities on the order of $500 - 700$ km/s. Assuming that these velocity flows trace the underlying gravitational potential wells and hence the underlying mass distribution (which is a more reasonable assumption than that requiring the visible matter to trace the dark matter), we are lead to conclude that there is substantial evidence for the dark matter of the universe to be distributed with substantial power on large scales.

The ambiguity in current structure formation scenarios is mostly an ambiguity in the choice of initial conditions. Choosing a WIMP dark matter candidate is essentially a choice of initial conditions for the spectrum and amplitude of the primeval density perturbations. We chose a standard CDM model with a Hubble constant of $H_0 = 75$ km/s Mpc$^{-1}$ and consider values of the density parameter $\Omega_0$ ranging from $\Omega_0 = 1$ to $\Omega_0 = 0.1$. Initial conditions are fixed by the Harrison-Zel'dovich spectrum and gaussian distributed amplitudes and phases for the initial scalar perturbations.

Our matter model is normalized using the top-hat $\sigma_8 = 1$ normalization:

$$\sigma_8^2 = \int_0^\infty 4\pi k^2 P(k) W^2(k) dk \tag{6}$$

where $P(k)$ is the Fourier transform of the square of the density perturbations $(\delta\rho/\rho)^2$ and

$$W(k) = \frac{3}{(kR_h)^3}(\sin(kR_h) - (kR_h)\cos(kR_h)) \tag{7}$$

is the Fourier transform of a spherical top hat of radius $R_h = 8h^{-1}$ Mpc. Using this normalization of the present $z = 0$ data, we linearly extrapolate back to the starting redshift of our simulation (typically $z = 100$) to obtain the initial amplitudes of our perturbations. Since perturbations on scales of $R_h \geq 8h^{-1}$ Mpc are still in the linear and quasi-linear regime, a linear extrapolation is justified. We use the same set of random seeds for the initial data throughout all of the simulations presented here. In this way we can make meaningful comparisons of the different simulations and contributions to the primary and secondary anisotropies.

An important quantity underlying the Rees-Sciama effect is the size and density contrast of the large scale void distribution in the universe. Even though we do not normalize directly to this distribution, we pay particular attention to both the typical and the largest voids generated by our numerical simulation. There is presently very little evidence for long ($\gg 100h^{-1}$Mpc) perturbations with significant amplitude in the universe. For example, Scaramella (1992) points out that "we obtain relatively low values for mass fluctuations on scales of wavelength $\lambda \approx 800$ Mpc, and an amplitude for expected CMB anisotropies consistent with present upper limits". Furthermore, Martinez-González, Sanz & Silk (1992) conclude that most of the secondary effect comes from scales where the spectrum has been observed ($< 180/h$ Mpc). In previous work (Anninos et al. 1991, TL95, Tuluie et al. 1995), typical void scales of $30h^{-1}$ Mpc were used in our simulations, with the largest voids about $60h^{-1}$ Mpc large. This is in good agreement with what has been observed (de Lapparent et al. 1986); the largest known void in Bootes is about $60h^{-1}$ Mpc big (Kirshner et al. 1981). For the Bootes void, Martinez-González& Sanz (1990) obtained a Rees-Sciama effect of $3 \times 10^{-7}$, in accord with our previous results (TL95). They conclude that for the empty regions observed in the local Universe the effect is, at maximum, of the order of $10^{-6}$. For the 'Great Attractor' using a radius of $100h^{-1}$ Mpc, they obtained a Rees-Sciama effect of $\sim 10^{-6}$. Blumenthal et al. (1992) use the reverse argument using COBE CMB results to place an upper limit of $80h^{-1}$ Mpc to the largest possible size of typical voids. However, to apriori rule out the existence of larger voids, say of typical size $\lambda \sim 100h^{-1}$ Mpc, may prove difficult if these voids exist with relatively low density contrasts. We have therefore allowed some of our simulations to explore domains in which $100h^{-1}$ Mpc voids are typical and the largest voids $\sim 200h^{-1}$ Mpc big. Even though observational evidence for such large voids is presently not at hand, we would like to consider their potential impact on the CMB via the Rees-Sciama effect. In particular, we are interested in whether the presence for larger voids will increase the Rees-Sciama effect if these voids are required to have density contrasts in accord with the scale invariant Harrison-Zel'dovich spectrum. Previous work has shown that for spherically symmetric and empty voids the Rees-Sciama effect increases with void size as $\propto \lambda^3$ (Panek 1992, see also Thompson & Vishniac 1987 and Meszaros 1994); however, the assumption of emptiness and special topologies of these voids, as well as the statistical effects of multiple voids and overdensities may well reduce this strong dependence on void size. The dependence of the Rees-Sciama effect, the proper motion effect and the decaying potential effect will be explored here as a function of structure size for realistic void distributions, topologies and density contrasts.

### 3.1. Numerical Technique for Matter Evolution

Our numerical study is based on a Particle-Mesh method (Hockney & Eastwood 1981, Anninos et al. 1991) in which the mass in the universe is dominantly in the form of dark matter. PM methods are particularly well-suited for collisionless systems of particles as they are designed to suppress two-body scattering. They are also fast and can handle many particles. The limited resolution, characteristic of PM methods due to the truncation of power on scales of order the mesh spacing, is not likely to affect our results significantly (as compared to using a $P^3M$ method) since the photon calculations are performed on a finite difference grid with the same truncation error as the PM grid on which the gravitational potential is solved.

The initial fluctuation spectrum of this collisionless matter is taken to be the Harrison-Zel'dovich spectrum (Harrison 1970, Zel'dovich 1972), modulated by damping processes that produce the CDM spectrum (Bardeen 1986). The initial positions of the particles at $z = 100$ are set up to be consistent with this density fluctuation spectrum. The particle positions are then evolved under the influence of gravity. Initially linear motion leads eventually to strong self-binding, and nonlinear condensations ($\delta\rho/\rho > 1$). We use the standard $\sigma_8 = 1$ normalization, Eqn. (6) in which the RMS overdensity in $8h^{-1}$ Mpc spheres is unity at $z = 0$. The final structure then resembles the familiar large scale sheet/filament/knot structures observed in the universe (Geller & Huchra 1986), including large voids.

We fix the maximum physical size of our Newtonian simulation for structure formation by requiring a sub-horizon size box during all epochs of the simulation. We therefore choose a maximum comoving box size of $360h^{-1}$ Mpc and a starting redshift of $z = 100$. The greatest limitation of our technique comes from the finite resolution of the computational grid on which the gravitational potential is defined. At present, a resolution of $7.5h^{-1}$ Mpc, corresponding to a single zone of the mesh is the limit for a $64^3$ simulation of the $360h^{-1}$ Mpc domain. At this resolution we will not be able to resolve the cluster collapse sufficiently. However, the $120h^{-1}$ Mpc domain does have adequate resolution ($1.9h^{-1}$ Mpc) to explore the Rees-Sciama effect from $\sim 20h^{-1}$ Mpc superclusters. On the other hand, a $120h^{-1}$ Mpc size simulation does not allow us to construct a distribution of large $\sim 100h^{-1}$ Mpc voids. Hence the need for different size simulation domains, $120h^{-1}$ Mpc to $360h^{-1}$ Mpc, all normalized equivalently.

### 3.2. Initialization

The dark matter perturbations are initialized on the grid with random phases and a gaussian distribution for the amplitude of each mode. We begin by placing particles uniformly distributed in the zone centers $\mathbf{x_0}$ and then calculate the displacements $\mathbf{p}(\mathbf{x_0})$ from that lattice, where $\mathbf{p}(\mathbf{x})$ is the inverse fourier transform of the power spectrum of perturbations $\mathbf{p}(\mathbf{k}) \propto \mathbf{k}\delta(\mathbf{k})/\mid \mathbf{k} \mid^2$.

In comoving coordinates, the particle positions are given by Zel'dovich's linear solution (Zel'dovich 1970):

$$\mathbf{x} = \mathbf{x_0} + \frac{b(t)}{a(t)}\mathbf{p}(\mathbf{x_0}), \tag{8}$$

where $\mathbf{x_0}$ is the fixed coordinate of the zone center and the second term describes the perturbation. The initial growing mode velocities $\mathbf{v}_g$ of the particles are then obtained by differentiating equation (8) and setting $d\mathbf{x_0}/dt = 0$ for a comoving grid

$$\mathbf{v}_g = \frac{d\mathbf{x_0}}{dt} + \mathbf{p}(\mathbf{x_0})\frac{d}{dt}\left(\frac{b}{a}\right) = \mathbf{p}(\mathbf{x_0})\frac{dz}{dt}\frac{d}{dz}\left(\frac{b}{a}\right), \tag{9}$$

where

$$\frac{dz}{dt} = -H_0(1+z)^2\sqrt{\Omega_0 z + 1}, \tag{10}$$

$$a(z) = (1+z)^{-1}, \tag{11}$$

$$b(z) = \sqrt{\Omega_0 z + 1}\,\Im(z), \tag{12}$$

$$\Im(z) = \int_z^\infty \frac{dz}{(1+z)^2(\Omega_0 z + 1)^{3/2}}. \tag{13}$$

We obtain

$$\mathbf{v}_g = -H_0(1+z)^2\mathbf{p}(\mathbf{x_0})\left\{(2 + 3\Omega_0 z + \Omega_0)\frac{\Im(z)}{2} - \frac{1}{(1+z)\sqrt{1+\Omega_0 z}}\right\}. \tag{14}$$

The integral $\Im(z)$ is evaluated for three separate cases of $\Omega_0$:

$$\Im(z) = \left(\frac{1}{(1-\Omega_0)(1+z)} + \frac{3\Omega_0}{(1-\Omega_0)^2}\right)\frac{1}{\sqrt{\Omega_0 z + 1}} + \frac{3\Omega_0}{2(1-\Omega_0)^2}\Im_\Omega(z), \tag{15}$$

where

$$\Im_\Omega(z) = \begin{cases} \frac{1}{\sqrt{1-\Omega_0}}\ln\left(\frac{\sqrt{\Omega_0 z + 1} - \sqrt{1-\Omega_0}}{\sqrt{\Omega_0 z + 1} + \sqrt{1-\Omega_0}}\right) & \text{for } \Omega_0 < 1.0, \\ \frac{2}{5}(1+z)^{-5/2} & \text{for } \Omega_0 = 1.0, \\ \frac{2}{\sqrt{\Omega_0-1}}\left(\arctan\frac{\sqrt{\Omega_0 z + 1}}{\sqrt{\Omega_0-1}} - \frac{\pi}{2}\right) & \text{for } \Omega_0 > 1.0 \end{cases} \tag{16}$$

For the case $\Omega_0 = 1.0$, $v_g$ reduces to

$$\mathbf{v}_g = \frac{2}{5}H_0\sqrt{1+z}\;\mathbf{p}(\mathbf{x_0}). \tag{17}$$

This now determines the initial velocities of the particles and together with the previously determined initial positions we can now evolve the matter. First, however, we need to normalize the initial amplitude of the fluctuations at the starting redshift $z_i$ of the simulation as described by Eqn. (6). For $\Omega_0 = 1$ the normalization is achieved by linearly extrapolating back the result of Eqn. (6) according to

$$\left.\frac{\delta\rho}{\rho}\right|_{z=z_i} = \left.\frac{\delta\rho}{\rho}\right|_{z=0}\frac{1}{1+z_i}. \tag{18}$$

For $\Omega_0 < 1$ one must use the growth law obtained from the equations of motion for hydrodynamical perturbations (Mukhanov et al. 1992):

$$\frac{\delta\rho}{\rho} \propto a\;\phi(a), \tag{19}$$

with $\phi(a)$ given by Eqn. (5). We find that the reduced growth of linear density perturbations in the $\Omega_0 = 0.3$ model is equivalent to a rescaling of $\delta\rho/\rho(z_i) = 2.2\;\delta\rho/\rho(0)/(1+z_i)$.

### 3.3. Evolution

The dynamical evolution of the dark matter is dependent only on the local values of the gravitational potential (and its gradient) for a given epoch characterized by its scale factor $a(t)$. The metric in the longitudal gauge is of the form

$$ds^2 = -(1+2\phi)dt^2 + a^2(t)\,(1-2\phi)\gamma_{ij}dx^i dx^j, \tag{20}$$

where

$$\gamma_{ij} = \delta_{ij}\left(1 + \frac{K}{4}\left(x^2 + y^2 + z^2\right)\right)^{-2}, \tag{21}$$

and $\delta_{ij} = 1$ if $i = j$ and zero otherwise. The constant $K$ takes the values -1, 0 and +1 for open, flat and closed universes respectively.

The equations of motion for the dark matter particles with subhorizon scale perturbations in the Newtonian limit $\phi \ll 1$ are

$$\frac{d\mathbf{x}}{dt} = \mathbf{v}, \tag{22}$$

$$\frac{d\mathbf{v}}{dt} = -2\frac{\dot{a}}{a}\mathbf{v} - \frac{\nabla_x \phi(\mathbf{x})}{a^2}. \tag{23}$$

Together with the Poisson equation for the gravitational potential,

$$\nabla^2 \phi = 4\pi G a^2 (\rho - \overline{\rho}), \tag{24}$$

where $\overline{\rho}$ is the proper background density, and the equation governing the evolution of the background scale factor

$$\dot{a}(t) = H_0 \sqrt{1 - \Omega_0 + \frac{\Omega_0}{a}}. \tag{25}$$

We begin our simulation at a redshift at which the numerical cube is at or below the horizon size. For $\Omega_0 = 0.3$ and our largest comoving simulation domain of $360h^{-1}$ Mpc the simulation box is 65% of the horizon at the maximum redshift of $z = 100$, while our smaller simulation boxes are a proportionally smaller fraction of the horizon size. The worst case is for our $\Omega_0 = 1.0$, $360h^{-1}$ Mpc box which is 18% above the horizon scale at $z = 100$. We verified that at this starting redshift all the simulations are still well within the linear regime for all scales and that no potential decay nor nonlinear structure formation has occurred yet.

As an additional check, we plot the average value of the absolute values of the potential perturbations across the entire grid as a function of redshift and compare to the analytic result. The good agreement between the theoretical evolution of $\phi$ and the result of our simulation for linear perturbations is shown in fig. 2. Late time nonlinearities in the density contrast $\delta\rho/\rho$ will modify the low $z$ region of the graph, but not leave nearly as significant an imprint in potential evolution as the decay of $\phi$ in the $\Omega_0 < 1$ model.

We use the $\sigma_8 = 1$ normalization described in section 3 for reasons of consistency with previous work. We find that this normalization yields a somewhat high large scale velocity flow of $600 - 900$ km/s averaged over the grid size. This is somewhat higher than the claimed $500 - 700$ km/s (Dressler et al. 1987, Bertschinger et al. 1990, Lynden Bell et al. 1988). Similarly, the RMS overdensity computed in $8h^{-1}$ spheres at $z = 0$ is up to 50 % higher than the linearly expected $\sigma_8 = 1$ value due to the accelerated growth of density perturbations at this quasilinear scale. The present day slope of the cluster-cluster correlation function is observed to be $\gamma = -1.8$ (Bahcall 1988). We find values of $\gamma \sim 1.7 - 2.1$. In summary, it seems that the standard $\sigma_8 = 1$ normalization yields somewhat higher values for our observables at $z = 0$, which is not unexpected since this normalization presupposes a linear growth rate of fluctuation and we instead follow matter fluctuations into quasi and nonlinear regimes.

## 4. Photon Simulation

We use a ray tracing technique (Anninos et al. 1991, TL95) to image the signatures of nonlinear matter structures in the CMB. Here we present a more detailed account of the modifications to the basic Anninos et al. 1991 technique.

A representative sample of photons are propagated back in time along their geodesics through the evolving dark matter structures, starting at the center of our computational box at $z = 0$ and finishing at $z = 100$. We propagate up to $10^5$ individual CMB photons, randomly spread over a chosen angular size square window along a chosen direction in the box, updating their individual temperature perturbations at every time step by computing explicitly the Rees-Sciama effect. Gravitational deflections (up to second order in the potential) by the matter structures are computed by solving explicitly the geodesic equation for every photon at every time step. We employed two numerically independent schemes for the computation of the Rees-Sciama effect. The first scheme directly integrates the comoving potential time derivative, second term in equation (1), for each photon along its geodesic. This comoving potential time derivative, $\partial\phi/\partial t$, is obtained directly from the matter simulation as follows: at every matter time step, the potential data and matter velocities are written to disk, then inverted and supplied to the photon simulation as input. The photon simulation computes $\partial\phi/\partial t$ as the difference in potential values for each grid point from matter step to matter step. In between the matter time steps, which are roughly a factor $(v_{matter}/c)^{-1}$ coarser than the photon time steps, $\partial\phi/\partial t$ is interpolated for each grid point to yield a smooth variation in time of the gravitational potential. As photons propagate through the grid, they accumulate successive values of $\partial\phi/\partial t$ at each photon time step. However, as photons traverse various grid points (one for each photon time step), they may sample substantially changing values of $\partial\phi/\partial t$ due to the spatial variations in $\phi$, even though $\partial\phi/\partial t$ itself is a slowly and smoothly varying function at each grid point. We therefore employ a volume weighted averaging of $\partial\phi/\partial t$ of the nearest neighbor grid points.

We have compared this technique against a second scheme which integrates the spatial potential gradients along the geodesic, that is the second term in equation (26):

$$\frac{\Delta T}{T} = \phi_e - \phi_r - 2 \int_r^e \frac{\mathbf{e} \cdot \nabla \phi}{a} dt, \tag{26}$$

where $\mathbf{e}$ is a unit vector along the photon's direction. Here the Rees-Sciama effect is obtained by subtracting the boundary terms off the integral term above. We found that the $\partial\phi/\partial t$ scheme is numerically more accurate and is used in all of our computations. The numerical errors produced by the spatial differentiation and interpolation, that are required in the second scheme, are in some cases an appreciable fraction of the Rees-Sciama signal.

### 4.1. Code Tests

As a first test we have constructed a transversely moving symmetric cluster potential of the form

$$\phi(r) = \frac{\phi_0}{r + r_0}. \tag{27}$$

In the center of mass frame of the moving cluster, the potential is static. The CMB proper motion temperature anisotropy of such a cluster moving in the transverse direction (see arrow) is shown in fig. 1. We find that our code reproduces the analytic result of Birkinshaw & Gull 1983 and Stebbins 1987. We further verified that the minimum and maximum temperature excursions scale proportional to the cluster's potential depth and transverse velocity, while the angular separation of the minimum and maximum temperature excursions is proportional to the angular size of the cluster's potential well.

Another important test of the validity of our approach is the requirement that boundary effects due to the periodicity of our computational domain are negligible. The largest angular scales we can explore are ones that will not extend beyond the size of the simulation box at $z = 100$. Otherwise we would image the repeated structures of multiple boxes. Furthermore, since photons travel through the simulation box several times due to our use of periodic boundary conditions, we must assure that no periodicity effect from repeatedly sampling the same symmetric direction are present. This is achieved by choosing a direction for our photon beam such that the photon beam traverses different regions each time it crosses the box. Obviously, symmetric directions, such as propagating the photons perpendicular to a cube side, are to be avoided for this reason. We test for periodicity effects by fourier transforming $\partial\phi/\partial t$ and checking that no significant frequencies are present. Since no outstanding features which would correspond to the length scale of the box are found in the fourier transform of $\partial\phi/\partial t$, we conclude that periodicity effects are properly handled by our method.

Our photon simulation relies on a finite number of photons to image the CMB temperature perturbations. The shot noise of this simulation depends on the particular physics we model. For example, in Tuluie et al. (1995), we model reionization and Thomson scattering, which tends to randomize the sky maps and distribute photons of greatly varying temperature perturbations into the same pixels of the sky maps, leading to a large shot noise. On the other hand, without Thomson scattering, we find that gravitational effects will only redistribute photons on arcminute scales, which is the pixel size, and hence the shot noise is small since now most of the photons in each pixel had similar histories. If we model the dependence of our results on the number of photons as

$$\frac{\Delta T}{T}(\infty) = \frac{\Delta T}{T}(N) + \alpha N^\gamma \tag{28}$$

where $\frac{\Delta T}{T}(\infty)$ is the extrapolated value for infinitely many photons (hence no shot noise), $\frac{\Delta T}{T}(N)$ is the actual value from the simulation and $N$ the number of photons. We performed five identical simulations with $N = (10^3, 5 \times 10^3, 10^4, 5 \times 10^4, 10^5)$ and found values of $\alpha = 4 \times 10^{-7}$ and $\gamma = -0.4$. This gives a truncation error of less than 1% for $10^5$ photons. For our results presented in this work we consistently use $10^5$ photons. For this few photons, a typical photon simulation takes only a few minutes of Cray C-90 CPU time.

### 4.2. Separation of Sources of Late-Time Gravitational Perturbations of the CMB

Our photon tracing technique provides a "total" contribution from secondary gravitational perturbations; that is, it models gravitational perturbations of the CMB from the intrinsic Rees-Sciama effect, the proper motion effect, and the decay of the gravitational perturbations due to $\Omega_0 < 1$. All three sources are effectively only represented by their influence on the individual values of $\partial \phi / \partial t$ on our computational grid. Hence further analysis is needed to separate their individual effects from the total skymap image obtained by our numerical scheme. This separation is useful in order to understand their relative importance.

Of course, any actual observation of the CMB anisotropies would only detect the sum total of these three effects plus the Sachs-Wolfe effect, Doppler shifts, foreground sources, etc. (see White et al. 1994 for a review). However, the point of separating the relative importance of these three sources of late-time gravitational perturbations of the CMB is that we want to ascertain a level of relevance for each source independently.

In particular, we wish to separate the temperature anisotropy resulting from the collapse and proper motion of the matter from the temperature anisotropy due to the decay of the gravitational perturbations in the $\Omega_0 < 1$ case. We accomplish this task by globally rescaling the gravitational potential at every grid point and at every matter time step according to figure 2 so that $\phi(z)_{avg}$ will be constant. The rescaled gravitational potential perturbations will not show any decaying effect and will only lead to the intrinsic and the proper motion Rees-Sciama effect. The results of the separation of the temperature anisotropy from the proper motion and intrinsic effect from the decaying perturbation effect is presented below.

### 5. Results

We have considered simulations with values of $\Omega_0 = (0.1, 0.3, 0.6, 1.0)$ in a $240h^{-1}$ Mpc cube. Additional simulations for $\Omega_0 = (0.3, 1.0)$ were carried out in $120h^{-1}$ Mpc and $360h^{-1}$ Mpc domains. Typical void sizes range from $\sim 30h^{-1}$ Mpc in the $120h^{-1}$ Mpc cube to $\sim 100h^{-1}$ Mpc in the $360h^{-1}$ Mpc simulation. Furthermore, in order to ascertain errors generated by our choice of a randomly generated initial seed, we have run a total of five matter simulations with different initial seeds for each value of $\Omega_0$ in the $240h^{-1}$ Mpc cube. The results are presented as follows: first, we discuss the total effect from secondary gravitational temperature anisotropies for simulations of increasing scale for the $\Omega_0 = 1.0$ model. Next we discuss simulations of increasing scale for the $\Omega_0 = 0.3$ simulations. Finally we discuss the dependence of secondary gravitational

CMB temperature perturbations on the value of $\Omega_0$ for a constant $240h^{-1}$ Mpc box size.

For $\Omega_0 = 1.0$, we find that secondary gravitational temperature anisotropies are generated during the nonlinear phase of structure formation only since $\phi =$constant during $\delta\rho/\rho \ll 1$. In particular, figure 3a shows how the total effect, comprised of the intrinsic Rees-Sciama effect and proper motion effect, is accumulated for our photon bundle as it propagates through our model universe. No monotonically increasing accumulation of the anisotropy is observed, yet complete cancelation between voids and overdensities does not occur either. Most of the secondary CMB temperature anisotropy is accumulated for redshifts less than $z \sim$ few, corresponding to the nonlinear epoch of structure formation. The decaying potential effect, which operates at larger redshifts, is not significant here - figure 3b shows that over 90 % of the total signal is comprised of the intrinsic Rees-Sciama and proper motion effects. This is true for any scale simulation, with little variation between various sized simulation domains. However, the levels of the temperature anisotropy differ as we allow the void and structure scale to increase with increasing box size. Typical values of $\Delta T/T \sim 3-4 \times 10^{-7}$, with a cosmic variance of about 15% for angular scales of a few degrees (TL95), are obtained for the $120h^{-1}$ Mpc box, where the void size is typically $\sim 30h^{-1}$ Mpc. We find that these voids contribute relatively little to the signal. Instead overdensities generate most of the signal. Even as we increase our typical void size to $\sim 100h^{-1}$ Mpc and maximum void size to $\sim 200h^{-1}$ Mpc ($360h^{-1}$ Mpc cube), we still find no complete domination of the signal by the voids and only a modest increase in $\Delta T/T$ to $\sim 7 \times 10^{-7}$ .

The weak signature of voids and stronger signature of overdensities is also supported by our skymaps: plate 1 shows the total effect in a region of our model universe located at redshift $z = 0.425$ and of thickness $\Delta z = 0.025$ in a $8° \times 8°$ window; plate 2 shows the corresponding density field, i.e. the underlying clusters, superclusters and voids in that region of our $240h^{-1}$ Mpc box. Even for this larger computational box, with correspondingly larger void sizes, the intrinsic Rees-Sciama and proper motion effects are still not dominated by the void signatures. Instead, the proper motion effect leaves a clear signature in the center of plate 1 and two weaker signatures in the bottom left corner of plate 1. A direct correlation to the clusters and super clusters in plate 2 is visible – the clusters of plate 2 are located at the center of the dipolar patterns of plate 1.

We conclude that as long as the void and large scale structure spectrum follows the scale invariant Harrison-Zel'dovich spectrum, the Rees-Sciama signal maintains an approximately constant value even as the typical void size is allowed to increase. Our argument in section 2.1, below Eqn. (2), supports this result. Figure 3a shows that above a $\sim 60h^{-1}$ Mpc typical void size (i.e. the $240h^{-1}$ Mpc cube) the intrinsic Rees-Sciama and proper motion effects level off - hence an upper limit to the effects exist even if we do not rule out the existence of very large ($\sim 100h^{-1}$ Mpc) voids. We conclude that the signal is at or below $\Delta T/T \sim 10^{-6}$ if $\Omega_0 = 1.0$.

Next we discuss the results for our $\Omega_0 = 0.3$ model. Here, we find an increased contribution from secondary gravitational effects. However, even though the increase is more substantial, the signal is still not as large as the primary Sachs-Wolfe effect. We find the largest signal ($\Delta T/T = 1.3 \times 10^{-6}$) occurs for $\Omega_0 = 0.3$ ($360h^{-1}$ Mpc box). A notable difference with the $\Omega_0 = 1.0$ results is the epoch at which the secondary CMB anisotropies are accumulated. From figure 4a one can see that now the accumulation of the secondary signal occurs much earlier. This is due to the decaying gravitational potential, which decays at about $z \sim 20$ (see fig.2), where a 10% reduction from the primordial potential has already occurred. Even thought the local potential perturbations change by a factor of order $10^{-5}$ between decoupling and now (fig. 2), CMB photons will not experience a secondary temperature anisotropy of that magnitude. Using a simple scaling argument, the magnitude of the decaying potential effect for a large scale potential perturbation near $z = 0$ is roughly

$$\frac{\Delta T}{T} \sim \dot\phi\, t_c \sim \phi \frac{l_c}{l_H} \sim 10^{-6} \frac{l_c}{100 h^{-1}\mathrm{Mpc}}, \tag{29}$$

where $t_c$ is the structure crossing time, $l_H$ the horizon length scale and $l_c = c\, t_c$ the size of the structure's potential perturbation region. Even for large structures, such as $100h^{-1}$ Mpc voids, the cumulative effects from several structures will be at the $10^{-6}$ level. However, the effect increases roughly linearly with structure size, a result also seen in figure 4a. The dominance of the decaying gravitational potentials is also born out in figure 4b, where now the signal coming from decaying potentials is about three times that of the sum of intrinsic Rees-Sciama and proper motion effects. The temperature plates show a similar result: plate 3 shows the total secondary signal, while plate 4 shows the potentials for that region of our simulation, i.e. the Sachs-Wolfe signal. The obvious negative correlation between these two maps indicates that the total secondary signal is directly related to the gravitational potentials. This is expected if the secondary signal is dominated by $\partial\phi/\partial t$ of the decaying gravitational potential. From Eqn. (1), a decaying potential term will enter with the opposite sign of the Sachs-Wolfe term. Hence, a high potential region (such as in a void) that is decaying will leave a blueshift for the Sachs-Wolfe effect and a redshift for the decaying potential effect, while a low potential region (such as near a cluster) will leave a redshift for the Sachs-Wolfe effect and a blueshift for the decaying potential effect. Thus, the decaying potential effect reduces the total CMB anisotropy somewhat.

On the other hand, we do not find a direct correlation to the underlying density field of the large scale structures any more. Plate 3 lacks the dipolar features of the $\Omega_0 = 1$ case caused by the proper motion effect. Instead, secondary temperature anisotropies are clearly dominated by the decaying potential perturbations as the strong negative correlation to the potential map (plate 4) shows. In summary, the Rees-Sciama signal from both the intrinsic evolution of structures and voids, as well as from the proper motion of the clusters, is dominated by the effect from the decaying gravitational potential wells in an $\Omega_0 < 1$ universe.

Last, we discuss the relative importance of nonlinear effects (the intrinsic Rees-Sciama effect and proper motion effect) with respect to the decaying gravitational potential effect as a function of $\Omega_0$. Figure 5 shows the dependence of the secondary gravitational signal on the value of $\Omega_0$, for a range of $\Omega_0 = 1.0$ to $\Omega_0 = 0.1$, where each point represents an average of five different matter simulations; each run executed with a different random initial seed. The results for temperature anisotropies were obtained by averaging identical photon simulations (identical except for the initial seeds in the matter runs) at each angular scale. Fig. 5 shows, that for decreasing values of $\Omega_0$, the rescaled potential simulations result in smaller and smaller intrinsic and proper motion Rees-Sciama effects, while the relative importance of the decaying potential effect increases. We find that for $\Omega_0 = 0.1$ as much as 80 % of the secondary temperature anisotropy is now due to the decaying potential effect, while for $\Omega_0 \sim 0.5$ both nonlinear (i.e. intrinsic evolution and proper motion effects) and the decaying potential effect are equally important. This is to be expected since in a low $\Omega_0$ universe the rate of clustering is reduced, leading to smaller intrinsic and proper motion Rees-Sciama effects, while the decaying potential effect increases in importance according to Eqn. (5).

## 6. Conclusion

We have investigated the imprint of secondary gravitational effects on the CMB. Specifically, we have considered the importance of the Rees-Sciama effect due to: (1) intrinsic changes in the gravitational potential of forming, nonlinear structures, (2) proper motion of nonlinear structures, and (3) late time decay of gravitational potential perturbations in open universes. Our numerical tool to investigate these effects offers certain advantages to the study of these nonlinear effects, but it prohibits an expansive parameter study; hence, we have selected several realizations of a CDM universe which can serve as indicators for a wider class of models. We can model the combined effect of the three sources of secondary gravitational effects on the CMB and weight their relative importance. We do not find a strong dependence of the total secondary temperature anisotropy on void size as long as the spectrum of matter fluctuations was Initially a Harrison-Zel'dovich one. Another advantage of our method is a statistical representation of the void, cluster and super cluster distribution that is carefully normalized to the actual observed matter structures and can, through the use of our ray tracing technique, be compared directly to the resulting primary CMB anisotropies. In addition, we model secondary gravitational effects on the CMB from the linear, quasi-linear and nonlinear regimes without any assumptions of their matching or scale dependence. Apart from a choice of dark matter and proper normalization of the matter model, no assumptions are made about the nature of CMB anisotropies: their scale dependence, structure topology and statistical distribution all arise naturally out of our model. The major disadvantage of our model is its limited dynamical range. As computational resources improve, this disadvantage can be reduced by performing higher resolution studies. Another disadvantage of our model is the requirement of sub-horizon size matter fluctuations. We are currently working on a generalization to superhorizon scales (Laguna & Anninos 1995).

In summary, our results for the importance of secondary, nonlinear gravitational effects on the CMB are as follows: For an $\Omega_0 = 1$ universe, the intrinsic Rees-Sciama effect and the effect from the proper motion of nonlinear structures are at a level $\leq 10^{-6}$. In particular, if we require void structures no larger than the largest observed void in Bootes, which is $60h^{-1}$ Mpc, then the effect drops to $3 - 4 \times 10^{-7}$. Even if typical large scale voids of $\sim 100h^{-1}$ Mpc and largest voids $\sim 200h^{-1}$ Mpc are present, the effect does not increase further with void size and levels off just below $10^{-6}$. As long as the void and large scale structure densities follow the scale invariant Harrison-Zel'dovich spectrum, the Rees-Sciama signal does not increase further even if the typical void size is allowed to increase. For $\Omega_0 < 1$, the intrinsic Rees-Sciama effect and the effect from the proper motion of nonlinear structures are not the dominant source of secondary temperature anisotropies from gravitational effects. Instead, the added secondary anisotropy due to the late time decay of gravitational potential perturbations in open universes dominates the secondary features and serves to decrease the total CMB anisotropy as it enters with the opposite sign of the Sachs-Wolfe anisotropy. For $\Omega_0 = 0.3$, secondary anisotropies reach a level of $\sim 1 - 2 \times 10^{-6}$ for $\sim 100h^{-1}$ Mpc voids, with a roughly linear dependence on void size.

The three sources of secondary CMB anisotropy considered in this study do not alter our previous understanding of the origin of the CMB anisotropies (Sachs & Wolfe 1967) greatly. These secondary temperature anisotropies are not a substantial part of the CMB sky perturbations. Maybe once CMB experiments achieve accuracies below the $10^{-6}$ level one could hope for a detection of these secondary signals. However, their subtraction from all other sources of CMB sky signals will be quite difficult still.

We thank R. Matzner, T. Padmanabhan, D. Spergel and A. Stebbins for discussions and helpful suggestions. Work supported in part by the NASA (at Los Alamos National Laboratory), NSF Young Investigator award PHY-9357219, and NSF grant PHY-9309834. Computations were carried out on a Cray 90 machine at the Pittsburg Supercomputing Center.

box size = $240h^{-1}$ Mpc

| $\Omega_0$ | RS+PM $\times 10^{-7}$ | DP $\times 10^{-7}$ | Total $\times 10^{-5}$ | $\theta$ |
|---|---|---|---|---|
| 0.1 | 1.1 | 4.7 | 0.8 | 0.5 |
| 0.1 | 1.3 | 6.1 | 1.3 | 1.0 |
| 0.1 | 1.5 | 6.7 | 2.2 | 2.0 |
| 0.3 | 1.9 | 2.9 | 0.9 | 0.5 |
| 0.3 | 2.7 | 4.0 | 1.2 | 1.0 |
| 0.3 | 2.7 | 5.0 | 2.4 | 2.0 |
| 0.6 | 2.8 | 1.5 | 0.4 | 0.5 |
| 0.6 | 3.8 | 1.6 | 0.9 | 1.0 |
| 0.6 | 5.7 | 1.9 | 1.6 | 2.0 |
| 1.0 | 4.3 | 0.3 | 1.3 | 0.5 |
| 1.0 | 5.2 | 0.3 | 2.5 | 1.0 |
| 1.0 | 5.6 | 0.1 | 2.3 | 2.0 |

Table 2: Rees-Sciama and proper motion (RS+PM)$_{rms}$ temperature anisotropy in units of $10^{-7}$, decaying potential (DP)$_{rms}$ temperature anisotropy in units of $10^{-7}$ and total (including Sachs-Wolfe) CMB temperature anisotropy in units of $10^{-5}$ for various angular scales $\theta$ (degrees).

---

# FIGURE CAPTIONS

**Fig.1** CMB temperature anisotropy from the proper motion of a cluster with potential $\phi(r) = \phi_\circ/(r + r_\circ)$. The dipolar pattern yields direct information about the clusters parameters: the minimum and maximum temperature excursions are indicative of the cluster potential depth and transverse velocity, while the angular separation of the minimum and maximum temperature excursions is a measure of the angular size of the cluster's potential well. The contour levels are drawn at intervals of $\Delta T/T = 2 \times 10^{-8}$, with the solid lines representing a blue (hot) shift and the dotted lines a redshift (cold).

**Fig.2** Decay of potential perturbations in the $\Omega_0 = 0.3$ model. Plotted are the result of our simulation and the corresponding analytic expression of eqn. 5. The potential decay occurs as early as $z \approx 20$, at which point a 10% reduction in the original amplitude of $\phi$ is reached. It is important to notice that by $z = 0$, $\phi$ has been reduced by a factor of two from its primordial value.

**Fig.3** (a) Accumulation of the total secondary CMB anisotropy for $\Omega_0 = 1.0$ in $1°$, $2°$ and $3°$ windows of the respective 120, 240 and $360h^{-1}$ Mpc cubes. The anisotropy increases whenever the photon's path encounters a high density region, increasing significantly during the nonlinear stage of the matter evolution. The source for the secondary anisotropy in this case is the intrinsic Rees-Sciama effect and the proper motion effect since, for $\Omega_0 = 1.0$, the potentials are constant in the linear regime.

(b) Same as in (a) but here we show that the decaying potential effect is indeed absent for $\Omega_0 = 1.0$. Shown are the total anisotropy and the result in which the gravitational potentials in the simulations were rescaled at every time step to remove the decaying potential effect. Since both curves are nearly identical, we conclude that for the $\Omega_0 = 1.0$ model the decaying potential effect is neglidible. The total secondary CMB temperature anisotropy is comprised of the intrinsic Rees-Sciama effect and the proper motion effect only for this case.

**Fig.4** (a) Same as in figure 3a but for $\Omega_0 = 0.3$. In this case, the epoch of significant accumulation occurs much earlier, starting at redshifts near $z \sim 20$. This matches with the decay of $\phi$ from fig. 2, which also initiates near $z \sim 20$. The major source for the secondary anisotropy in this case is the decay of potential perturbations associated with open models.

(b) Same as in (a) but now the effect of the decaying potentials is shown. A clear separation is visible between the total effect and the intrinsic Rees-Sciama and proper motion effects. Both the Rees-Sciama and proper motion effects have been reduced due to the reduced growth of density perturbations in the $\Omega_0 = 0.3$ model, while the decaying gravitational potentials now provide an equally important contribution to the CMB anisotropy.

**Fig.5** Separation of the intrinsic Rees-Sciama and proper motion effects from the decaying potential effect as a function of $\Omega_0$. Shown are the average values for simulations with five different initial random seeds for each value of $\Omega_0$ and angular scale. The upper set of curves correspond to the total secondary temperature anisotropy and includes the intrinsic Rees-Sciama effect, the proper motion effect and the decaying potential effect. The lower set of curves has the decaying potential effect removed.

## PLATE CAPTIONS

**Plate.1** Temperature map of $10^5$ photons of the simulated microwave sky for a $8° \times 8°$ window of our $240 h^{-1}$ Mpc simulation, showing the total secondary temperature anisotropy from gravitational effects only. This plate shows the signal from an isolated region of the $\Omega_0 = 1$ model located at a redshift $z = 0.425$ and $\Delta z = 0.025$ thick, containing several typical clusters and voids of our model. Here the secondary temperature anisotropy is due to the proper motion and intrinsic Rees-Sciama effects. For this and subsequent maps the contour levels are adjusted to the minimum and maximum of each plate. Plate 1 has (min,max) levels of $(-4.3 \times 10^{-7}, 4.5 \times 10^{-7})$.

**Plate.2** The voids and clusters for the same region as plate 1. Here we integrated the density perturbations along the same photon histories as in plate 1 and for the same redshift interval. Notice that the clusters and super clusters lie at the center of the dipolar features of plate 1 - this is the proper motion effect from clusters. The signature of the $\sim 60 h^{-1}$ Mpc voids is small.

**Plate.3** Temperature map for the same direction as plates 1 and 2 in the $\Omega_0 = 0.3$ model in a $\Delta z = 0.03$ thick region located at a redshift $z = 0.33$ of our $240 h^{-1}$ Mpc box. For $\Omega_0 = 0.3$ the secondary temperature anisotropies are now dominated by the decaying potential effect. No clear dipolar features due to the proper motion effect as in plate 1 are observed. Plate 3 has (min,max) levels of $(-5.9 \times 10^{-7}, 10.4 \times 10^{-7})$.

**Plate.4** The gravitational potentials underlying the region of plate 3. Notice the large negative correlation with plate 3, which shows the origin of the decaying potential anisotropy is indeed the potential itself. Here we find $\Delta T / T_{rms} = 0.9 \times 10^{-5}$, corresponding to the primary anisotropy of the Sachs-Wolfe effect.

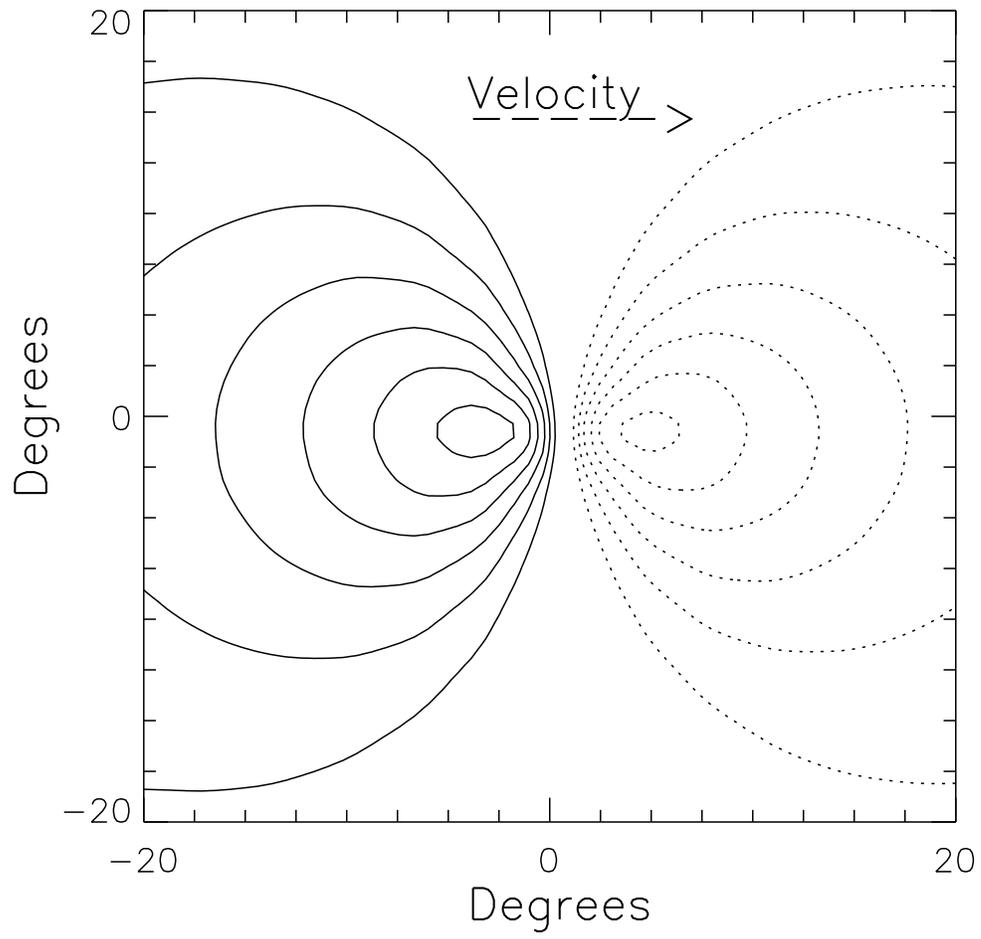

Fig. 1

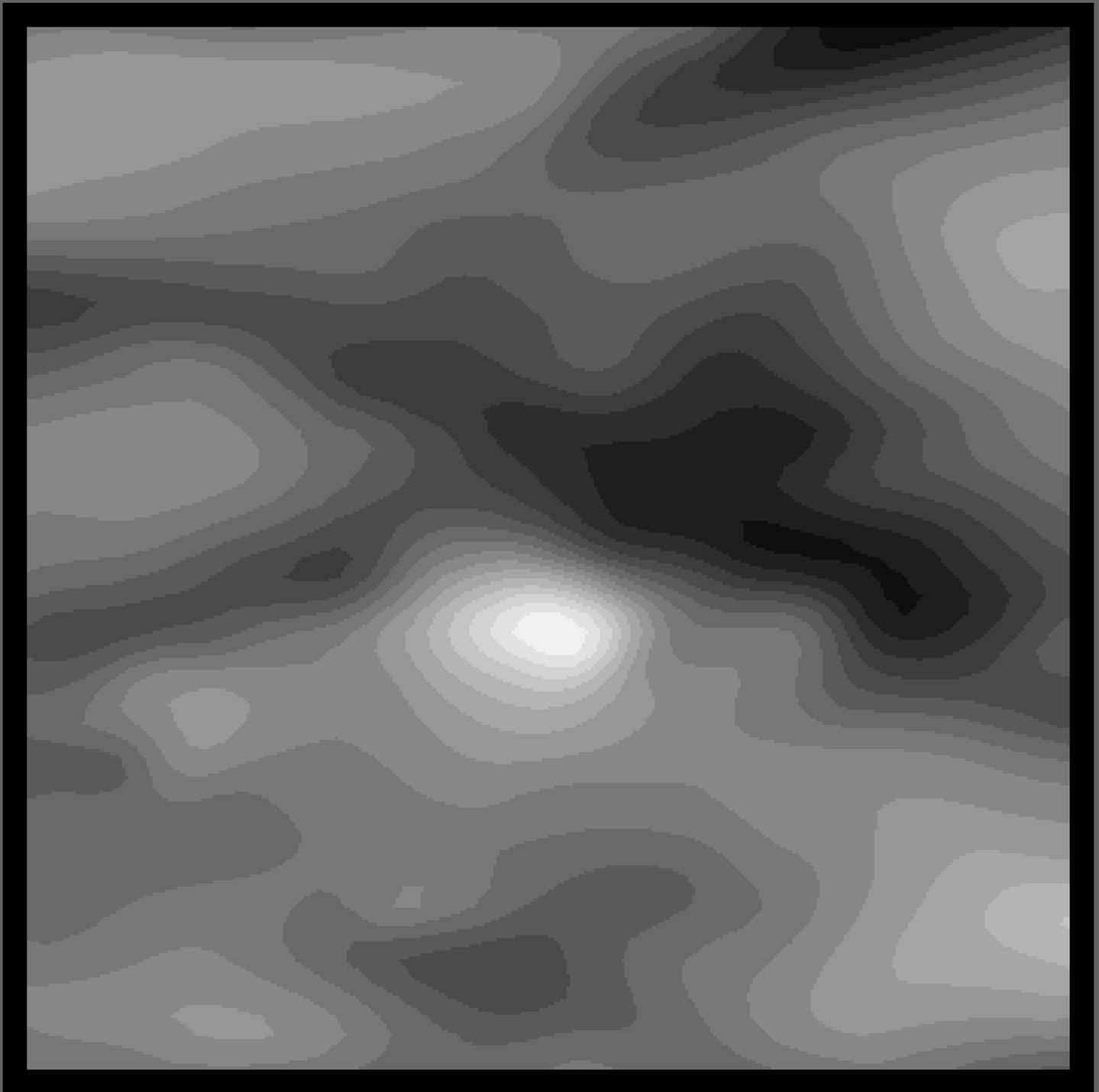

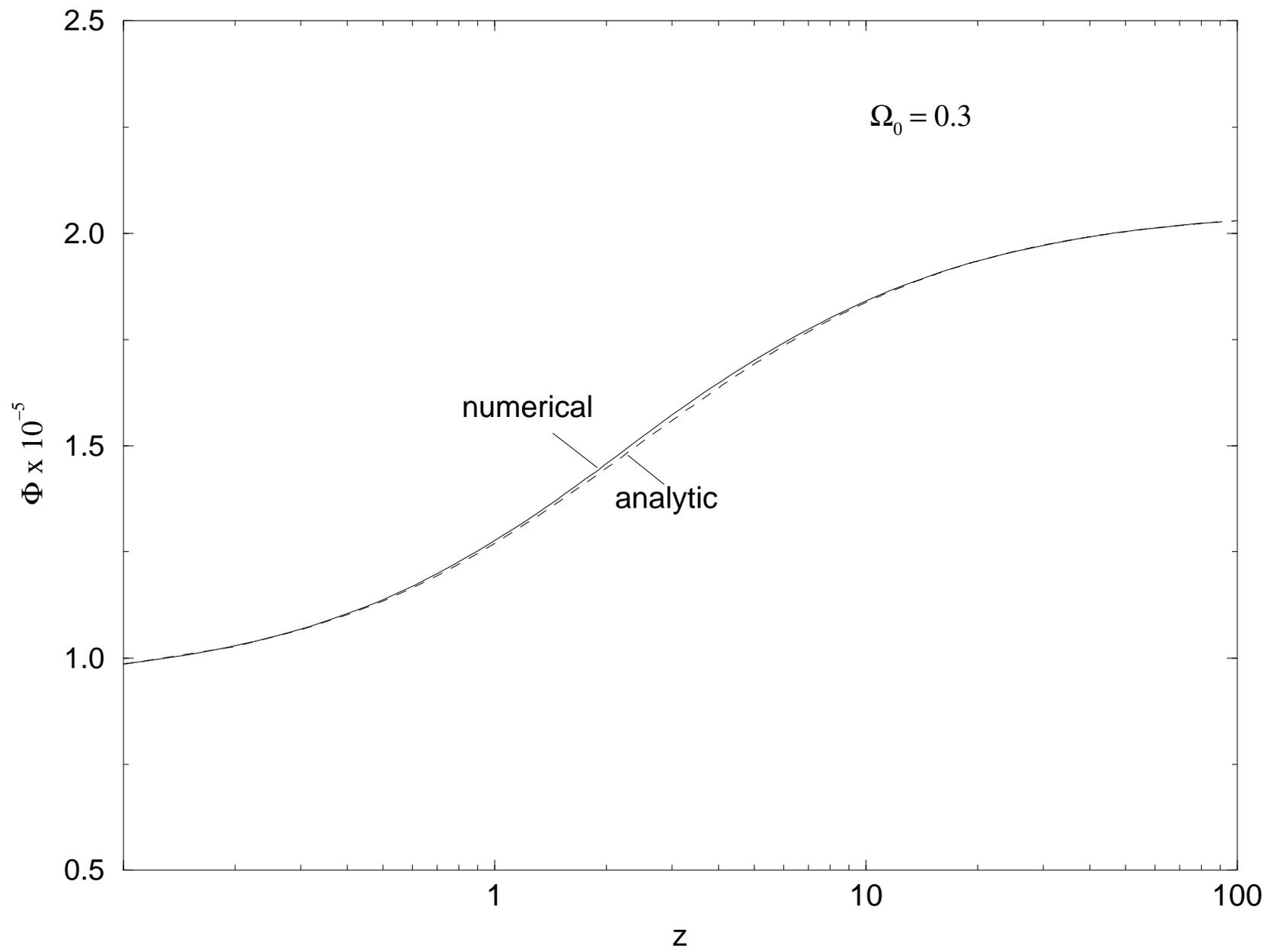

Fig. 2

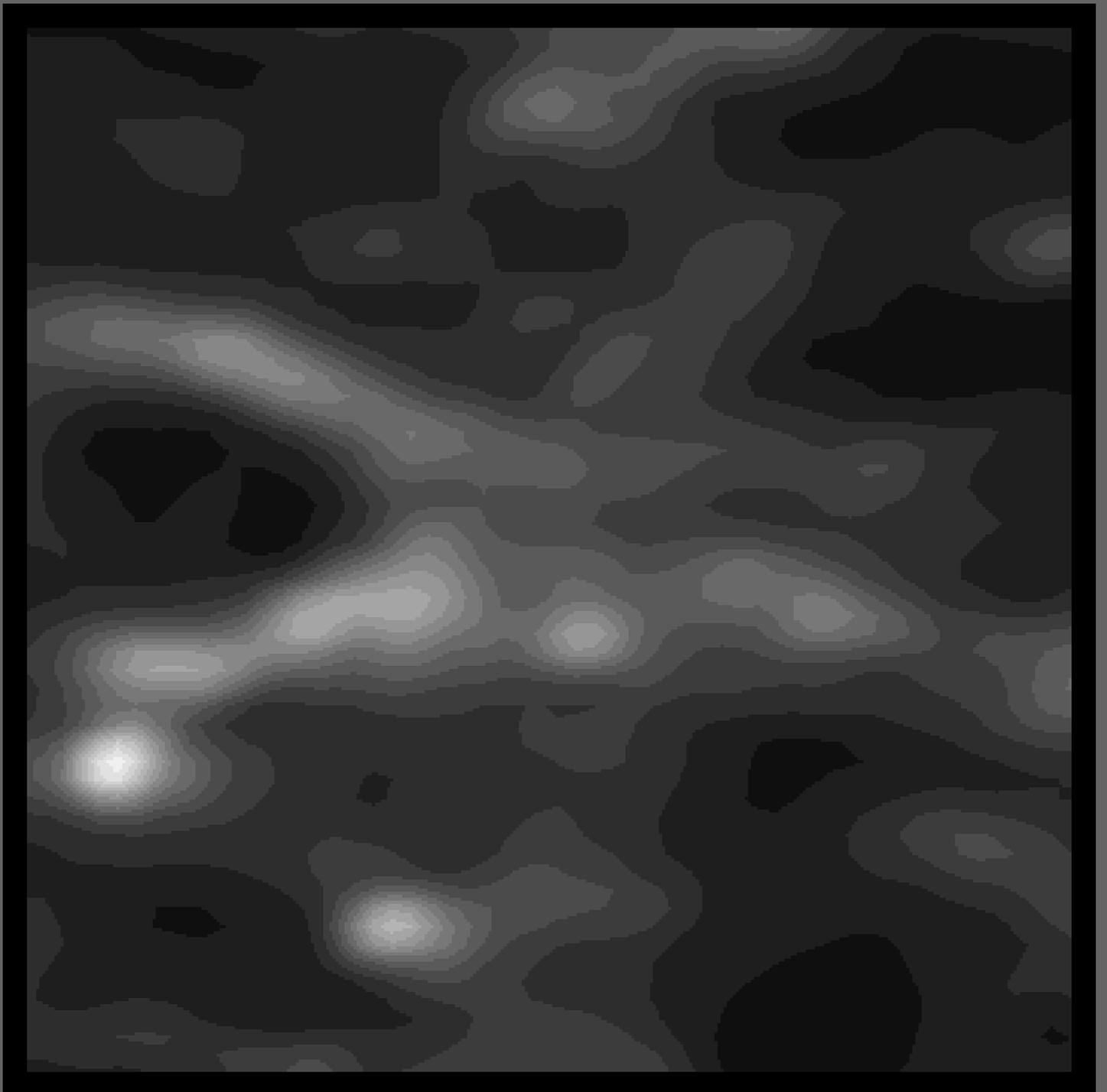

Fig. 3

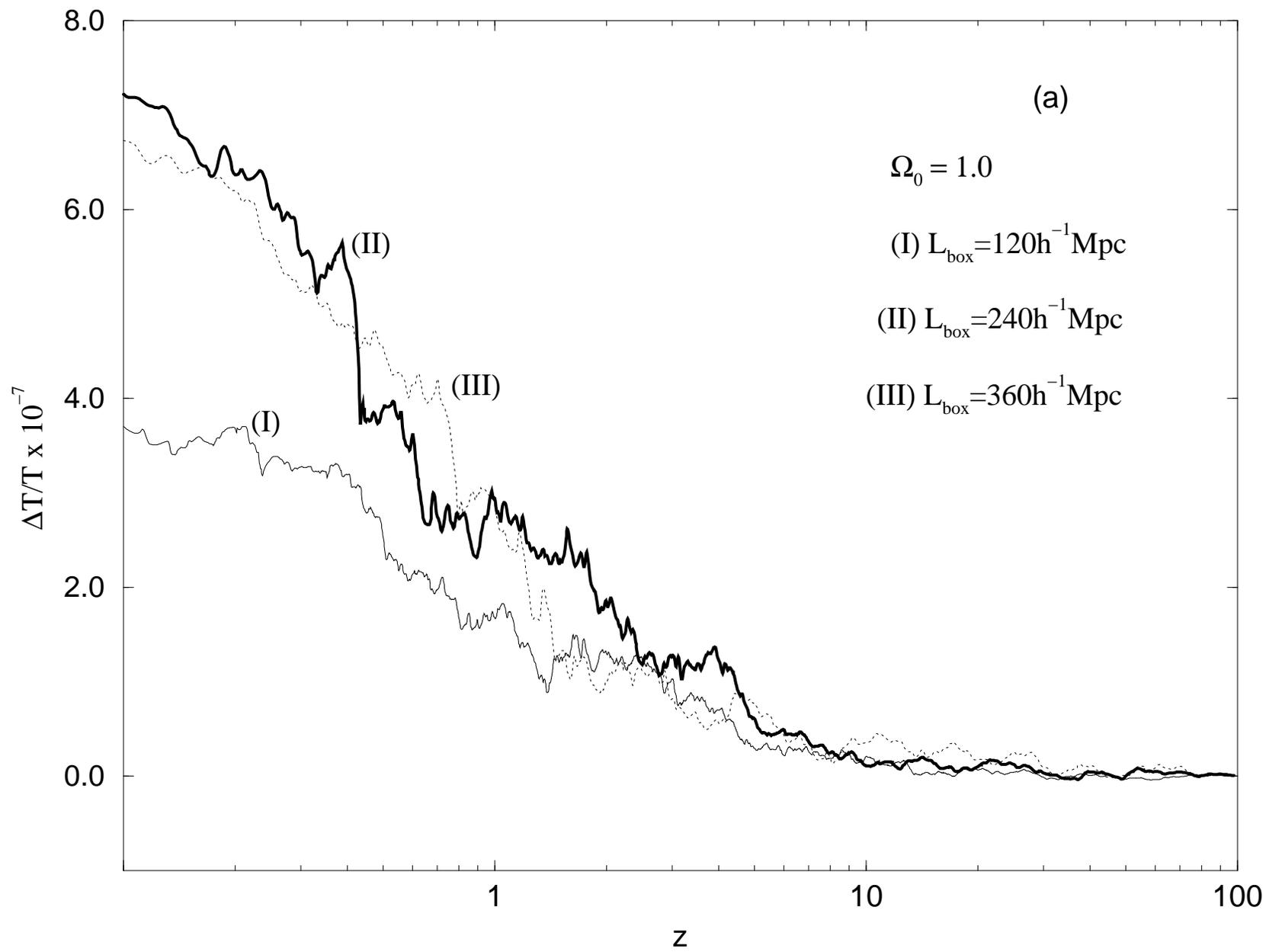

Fig. 3

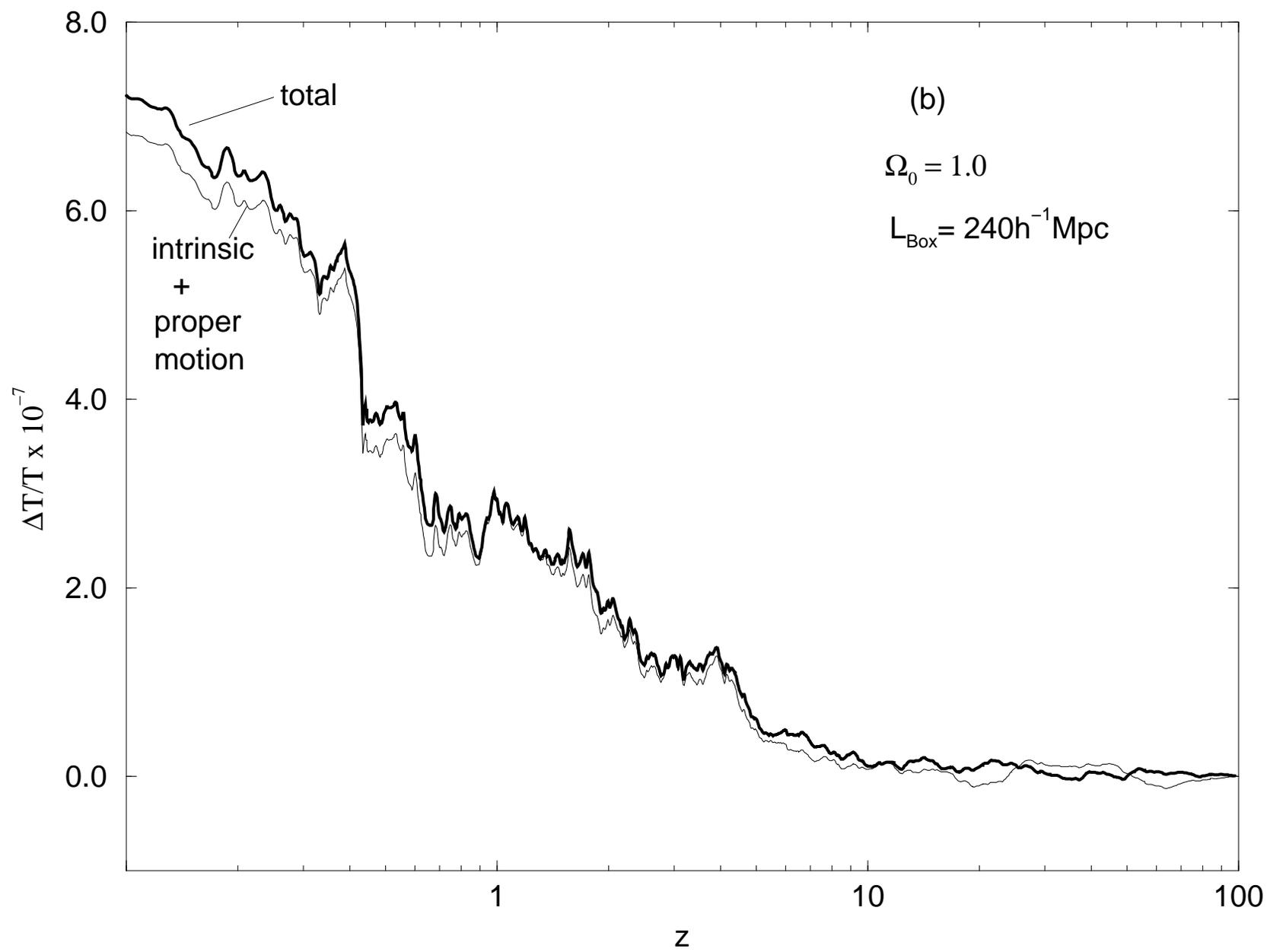

Plate.1

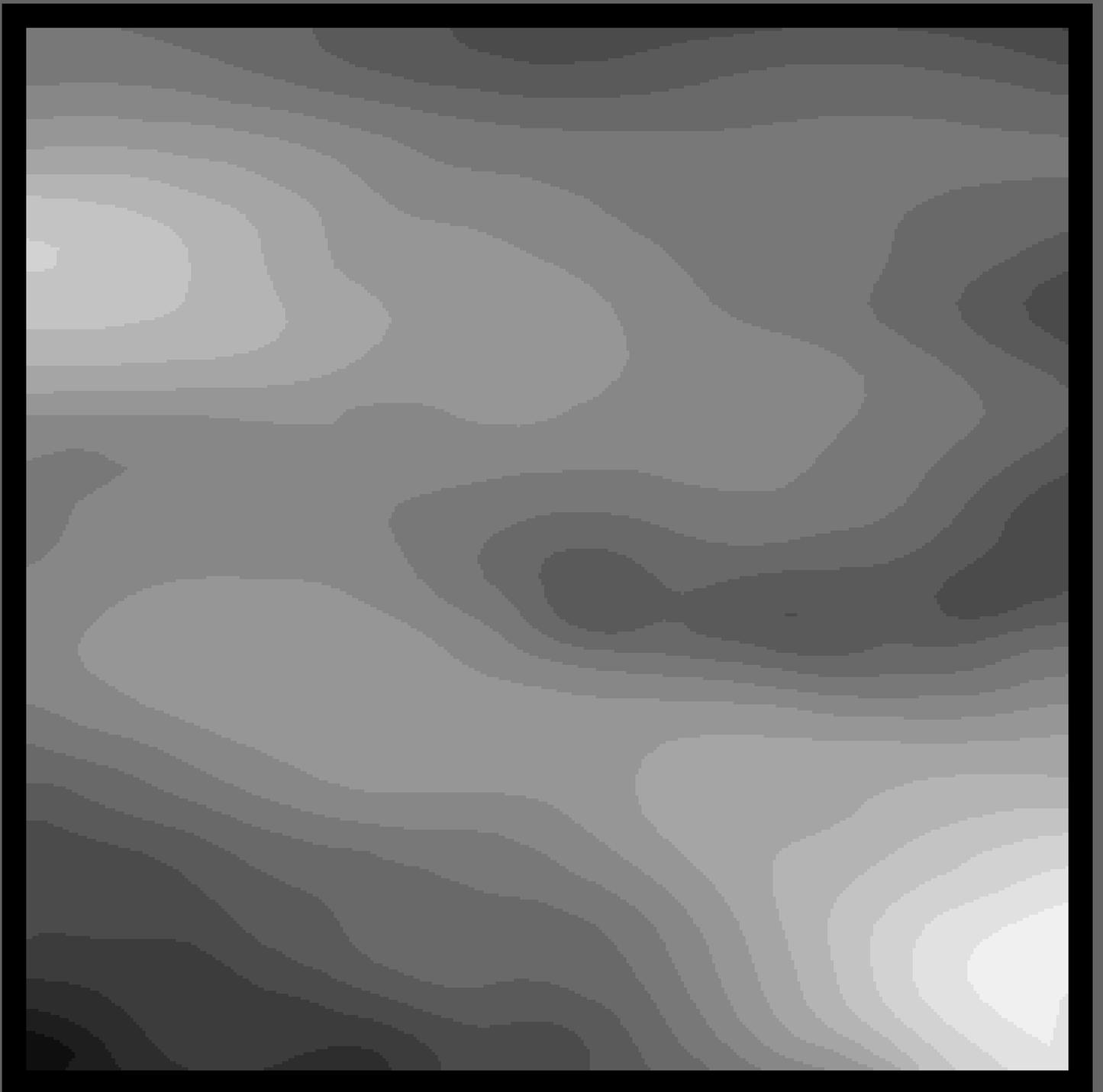

Fig. 4

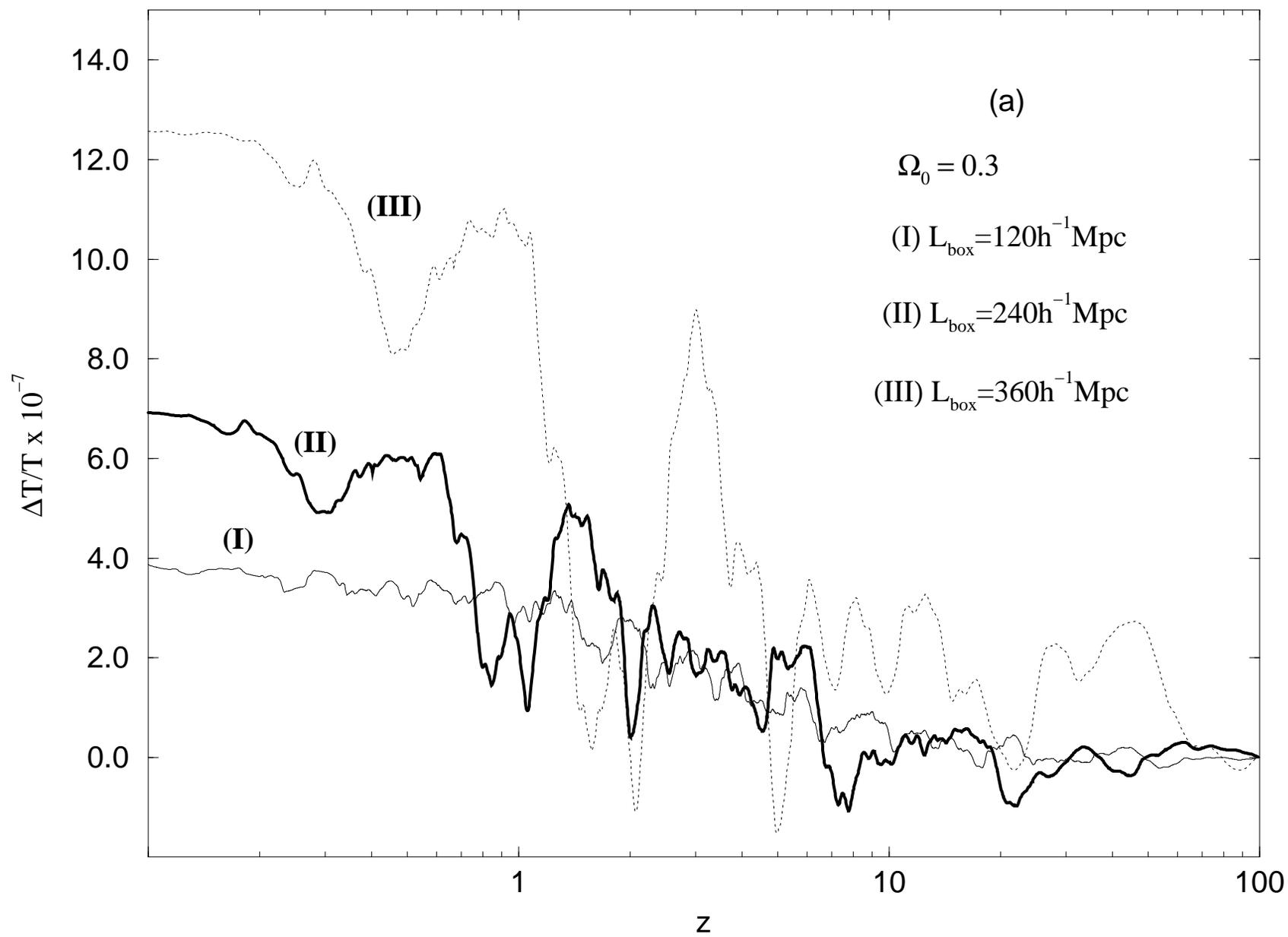

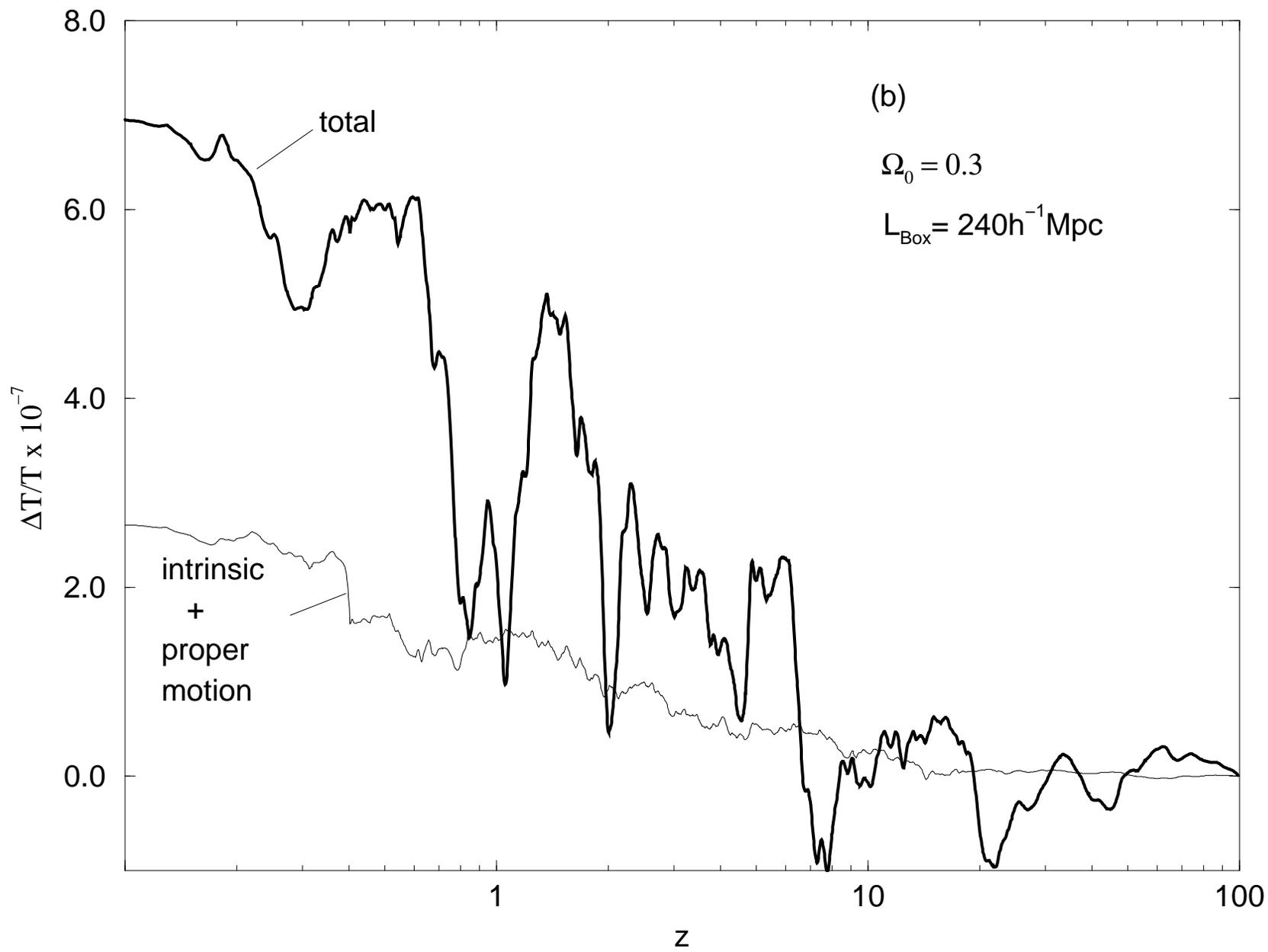

Fig. 4

Plate.2

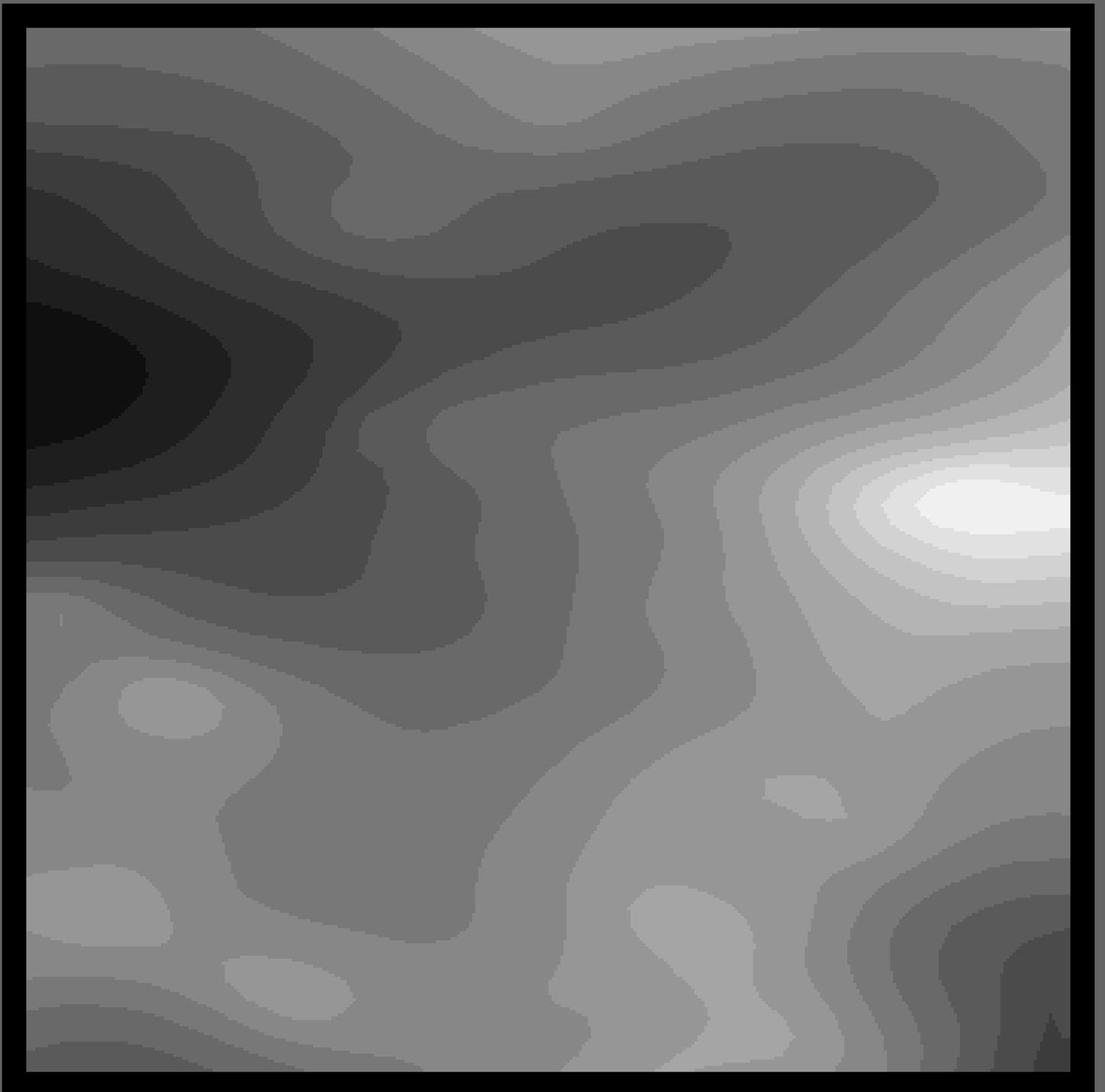

Fig. 5

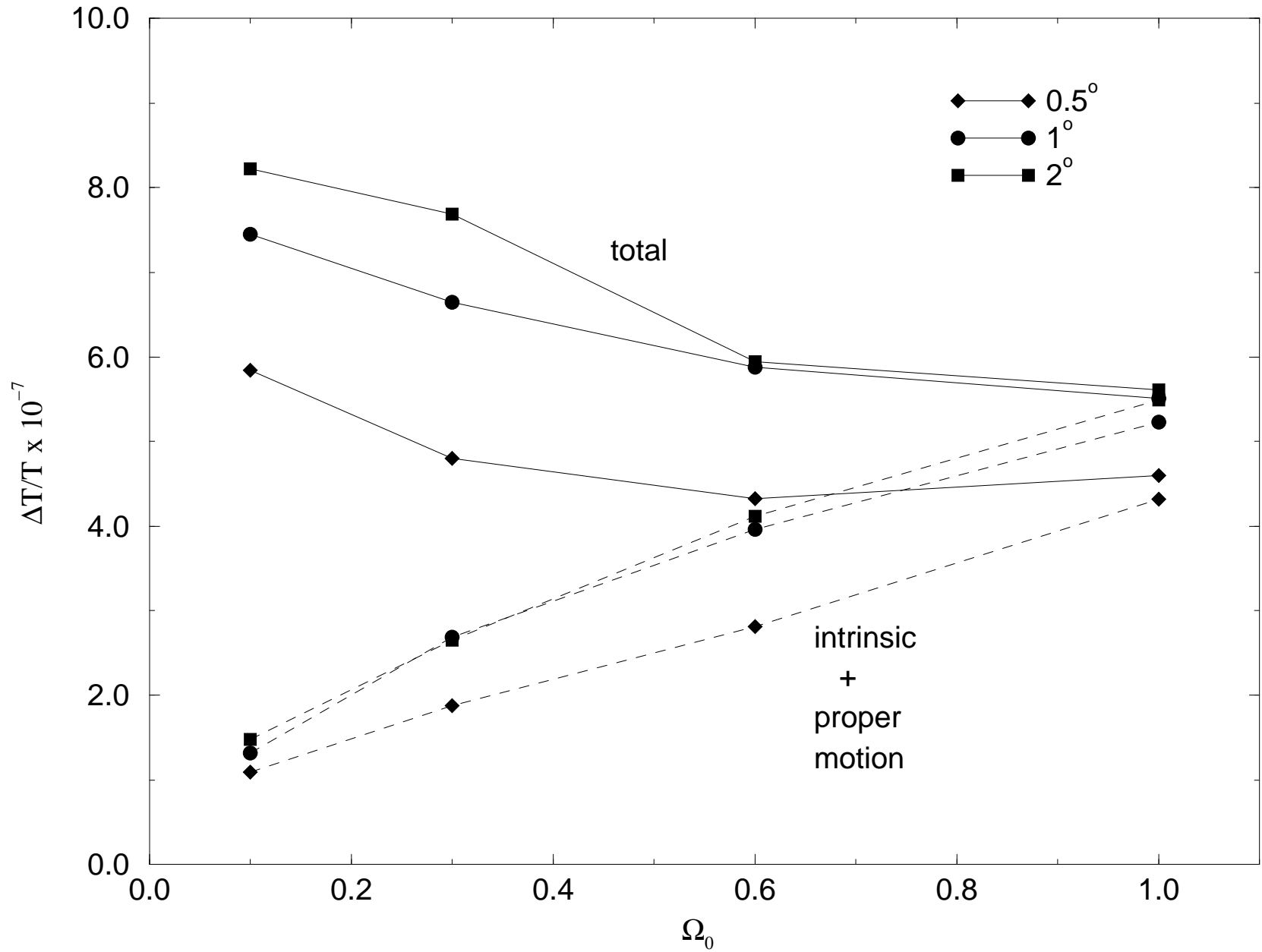